\documentclass{WileyMSP-template}

\begin{document}

\pagestyle{fancy}

\title{Radiative control of localized excitons at room temperature with an ultracompact tip-enhanced plasmonic nano-cavity}

\maketitle


\author{Hyeongwoo Lee$^\dagger$}
\author{Inki Kim$^\dagger$}
\author{Chulho Park}
\author{Mingu Kang}
\author{Jungho Mun}
\author{Yeseul Kim}
\author{Markus B. Raschke}
\author{Mun Seok Jeong}
\author{Junsuk Rho$^\ast$}
\author{Kyoung-Duck Park$^\ast$}

\begin{affiliations}
	
H. Lee, M. Kang, Prof. K.-D. Park\\
Department of Physics, Ulsan National Institute of Science and Technology (UNIST), Ulsan 44919, Republic of Korea\\
Email Address: kdpark@unist.ac.kr

I. Kim, Y. Kim, Prof. J. Rho\\
Department of Mechanical Engineering, Pohang University of Science and Technology (POSTECH), Pohang 37673, Republic of Korea

J. Mun, Prof. J. Rho\\
Department of Chemical Engineering, Pohang University of Science and Technology (POSTECH), Pohang 37673, Republic of Korea

C. Park, Prof. M.S. Jeong\\
Department of Energy Science, Sungkyunkwan University (SKKU), Suwon 16419, Republic of Korea

Prof. M.B. Raschke\\
Department of Physics, Department of Chemistry, and JILA, University of Colorado, Boulder, CO 80309, USA

\end{affiliations}


\keywords{Single quantum emitter, localized exciton, tip-enhanced photoluminescence, plasmonic structures, nano-cavity, Purcell effect}

\begin{abstract}

In atomically thin semiconductors, localized exciton (X$_{L}$) coupled to light shows single quantum emitting behaviors through radiative relaxation processes providing a new class of optical sources for potential applications in quantum communication. 
In most studies, however, X$_{L}$ photoluminescence (PL) from crystal defects has mainly been observed in cryogenic conditions because of their sub-wavelength emission region and low quantum yield at room temperature. 
Furthermore, engineering the radiative relaxation properties, e.g., emission region, intensity, and energy, remained challenging. 
Here, we present a plasmonic antenna with a triple-sharp-tips geometry to induce and control the X$_{L}$ emission of a WSe$_{2}$ monolayer (ML) at room temperature. 
By placing a ML crystal on the two sharp Au tips in a bowtie antenna fabricated through cascade domino lithography with a radius of curvature of $<$1 nm, we effectively induce tensile strain in the nanoscale region to create robust X$_{L}$ states. 
An Au tip with tip-enhanced photoluminescence (TEPL) spectroscopy is then added to the strained region to probe and control the X$_{L}$ emission \cite{lee2020}. 
With TEPL enhancement of X$_{L}$ as high as $\sim$$10^{6}$ in the triple-sharp-tips device, experimental results demonstrate the controllable X$_{L}$ emission in $<$30 nm area with a PL energy shift up to 40 meV, resolved by tip-enhanced PL and Raman imaging with $<$15 nm spatial resolution. 
Our approach provides a systematic way to control localized quantum light in 2D semiconductors offering new strategies for active quantum nano-optical devices.

\end{abstract}


\section{Introduction}

Ever since photoluminesence (PL) responses of localized exciton (X$_{L}$) were observed in naturally occurring crystal defects and edges in atomically thin semiconductors \cite{he2016, he2015, dang2020, zhang2017, srivastava2015, koperski2015}, their unique single quantum emitting behaviors in a fascinating 2D platform have attracted a lot of attentions for their potential applications in quantum information and communication devices \cite{duan2001, ou2008, shahriar2012}. In addition, recent studies have demonstrated that the X$_{L}$ can be deterministically induced at any desired locations. For instance, the X$_{L}$ emission was observed by transferring atomically thin semiconductors on nano-structures through the induced artificial strain \cite{iff2018, branny2017, palacios2017}. Furthermore, a nanocube cavity array was exploited to induce strain and drive more efficient X$_{L}$ emission through Purcell enhancement \cite{luo2018}, yet without the ability to control their radiative emission properties. As further practical applications, experiments inducing X$_{L}$ with nano-structures and controlling its frequency through actuators were reported \cite{chakraborty2020, kim2019}.

However, these studies have only been performed at cryogenic temperatures because of the small exciton binding energy of X$_{L}$ \cite{huang2016}, which leads to a much lower quantum yield at room temperature compared to the overwhelming radiative emission of neutral exciton (X$_{0}$) \cite{pandey2019, ross2013}. Since these low temperature operations heavily restrict the practical application of single quantum emitting devices, there have been significant efforts to observe the X$_{L}$ emission at room temperature. To probe the weak X$_{L}$ emission at room temperature, local near-field PL measurements are the only possible approach up to the present. For instance, the defect bound excitons in a WS$_{2}$ monolayer were spatially resolved using a near-field scanning optical microscopy (NSOM) with $\sim$70 nm spatial resolution \cite{lee2017}. In this work, it was found that the X$_{L}$ emission properties are strongly modulated by the structural features of the crystal defects. In addition, the strain-induced X$_{L}$ emissions at nanobubbles were probed using a scattering type NSOM approach \cite{darlington2020}. Through higher spatial resolution ($\sim$34 nm) nano-imaging and -spectroscopy, they revealed the correlation between the X$_{L}$ emission properties and the strain-induced confinement potential distribution with the support of a theoretical model. However, these near-field probing methods have low optical sensitivity and cannot artificially induce and control the X$_{L}$ emission. Therefore, it is highly desirable to deterministically induce, probe, and control the X$_{L}$ emission with high optical sensitivity and nanoscale spatial resolution at room temperature.

Here, we present a dynamic plasmonic nano-cavity combined with tip-enhanced nano-spectroscopy to deterministically induce, locally probe, and systematically control the X$_{L}$ emission at room temperature. We facilitate the localized states of excitons in a WSe$_{2}$ monolayer by inducing a tensile strain on the nanoscale using extremely sharp bowtie tips \cite{kim2020}. We then form the triple-sharp-tips plasmonic cavity by adding an Au tip on the bowtie nano-gap and measuring tip-enhanced photoluminescence (TEPL) of the X$_{L}$ with spatial resolution of $<$15 nm. In this case, the radiative emission rate of the X$_{L}$ is highly enhanced by the Purcell effect in the ultracompact cavity \cite{purcell1946, jimenez2020, akselrod2014} and we achieve $\sim$10$^{6}$-fold TEPL enhancement. In addition, we can dynamically tune the cavity mode volume and the corresponding excitation field strength and emission rate through the atomic force control of the Au tip with $\sim$0.2 nm precision \cite{park2019}, which enables the radiative control of X$_{L}$.

\section{Results}

\subsection{Triple-sharp-tips geometry for deterministic localized exciton emission}

\begin{figure}
	\begin{center}
		\includegraphics[width=18 cm]{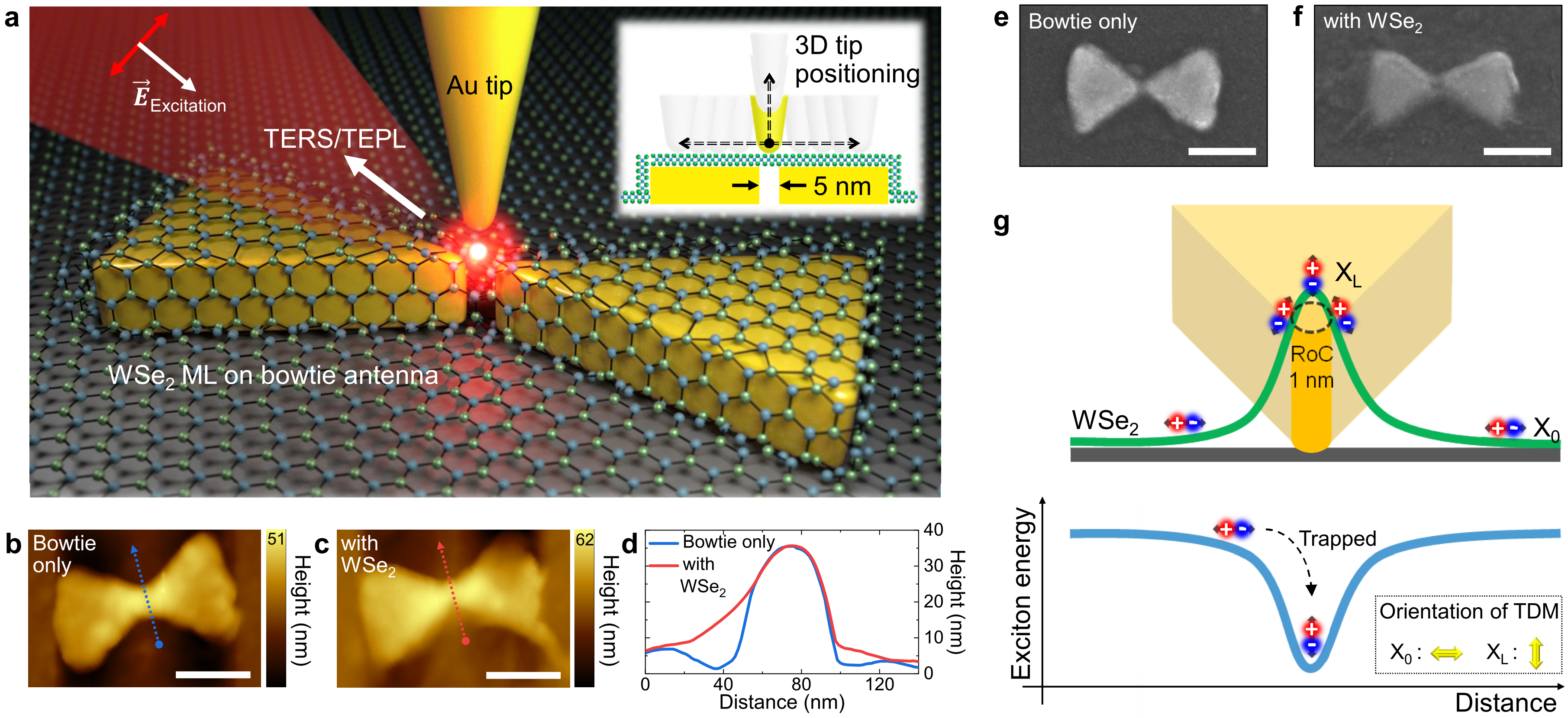}
		\caption{Experimental schematic and energy diagram of the triple-sharp-tips antenna with a WSe$_{2}$ monolayer.
			(a) Schematic of tip-enhanced nano-spectroscopy and triple-sharp-tips structure to induce, probe, and control the radiative emissions of localized excitons in a WSe$_{2}$ monolayer. The Au tip position can be controlled three-dimensionally through shear-force AFM feedback.
			AFM images of the bowtie nano-antenna without (b) and with (c) a WSe$_{2}$ monolayer and corresponding AFM line profiles (d) across the bowtie nano-gap derived from (b, blue) and (c, red). 
			SEM images of the bowtie nano-antenna without (e) and with (f) a WSe$_{2}$ monolayer. The transferred crystal clings tight on the sharp nano-gap. Scale bars are 100 nm.
			(g) Detailed schematic of the sharply curved WSe$_2$ crystal at the nano-gap and the distribution of X$_{0}$ and X$_{L}$ with corresponding energy diagram near the localized confinement potential. TDM, transition dipole moment.}
		\label{fig:fig1}
	\end{center}
\end{figure}

To induce artificial nanoscale strain in WSe$_{2}$ monolayer (ML), we use sharp tips in an Au bowtie structure fabricated by cascade domino lithography, which overcomes long-standing challenge of sub-nanometer tip fabrication \cite{kim2020}. The tips in the bowtie structure then has a 5 nm gap with a radius of curvature (RoC) of $<$1 nm.
When the WSe$_{2}$ ML is transferred to this extremely sharp structure, the locally induced strain forms a potential well, which can bound excitons tightly on the nanoscale \cite{hidouri2016, feierabend2019, branny2017}. To couple optical fields effectively to these X$_{L}$ and detect their PL responses, we use TEPL spectroscopy, as illustrated in \textbf{Figure~\ref{fig:fig1}}a \cite{park2018}. The electrochemically etched Au tip is positioned on the nano-gap to form the triple-sharp-tips geometry and able to control the distance between tip and sample with $<$0.2 nm precision using shear-force atomic force microscopy (AFM) feedback (see Experimental Section for details).

When a linearly polarized excitation beam (632.8 nm), polarized parallel with respect to the tip axis, is focused onto the triple-sharp-tips structure, both in-plane and out-of-plane optical fields are strongly localized which enable us to measure TEPL response of X$_{0}$ and X$_{L}$. We confirm the capping condition of the transferred crystal on the nano-gap through AFM and scanning electron microscopy (SEM) measurements, as shown in Figure~\ref{fig:fig1}b-f. As can be seen in the high-resolution 2D images and topographic line profiles, the ML crystal uniformly covers the bowtie structure and forms a stiff curve at the nano-gap. Figure~\ref{fig:fig1}g (top) shows an illustration of the steeply curved WSe$_{2}$ ML formed by the extremely sharp tips of the bowtie antenna. Since the strain is induced in the nanoscale steep curve, neutral excitons drift and are bound to the confinement potential as a form of the localized state \cite{darlington2020}. Specifically, a tensile strain is strongly induced into the vertical axis which decreases the energy state of the X$_{L}$ as described in Figure~\ref{fig:fig1}g (bottom).

\subsection{Nano-chemical characterization of the structure of transferred WSe$_{2}$ monolayer}

\begin{figure}
	\begin{center}
		\includegraphics[width=16 cm]{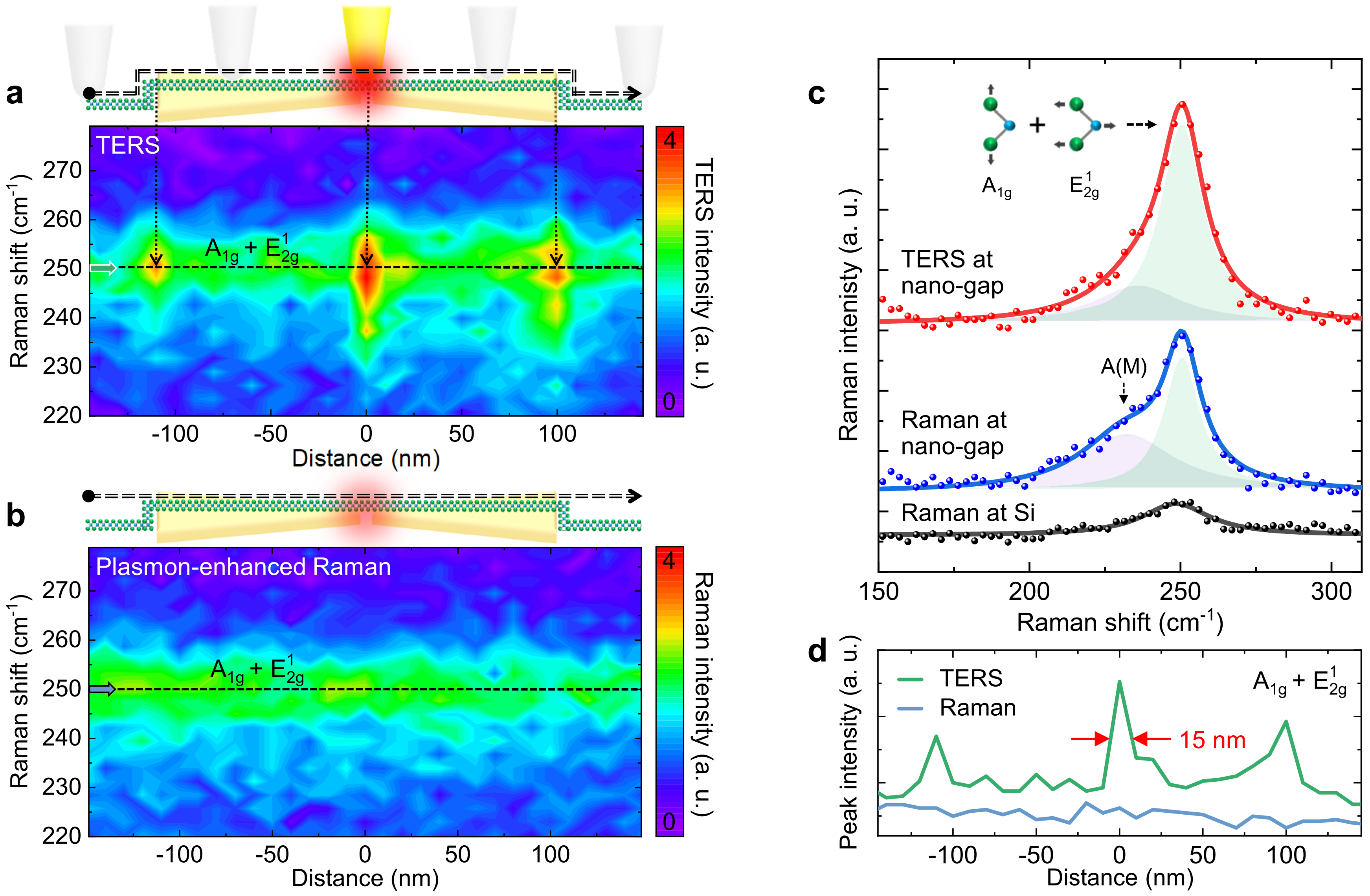}
		\caption{Probing the effect of sloping down regions of a WSe$_{2}$ monolayer with TERS measurement.
			Spectral TERS (a) and plasmon-enhanced Raman (b) line traces along the longitudinal center axis of the bowtie antenna with a WSe$_{2}$ monolayer measured at room temperature.
			(c) TERS (red) and plasmon-enhanced Raman (blue) spectra of a WSe$_{2}$ monolayer at the nano-gap with Voigt fitting, exhibiting a significant enhancement of degenerated A$_{1g}$ and E$_{2g}^{1}$ mode in the TERS spectrum. Far-field Raman spectrum of a WSe$_2$ monolayer on the Si substrate is shown for comparison.
			(d) TERS (green) and plasmon-enhanced Raman (blue) intensity profiles of the degenerated A$_{1g}$ + E$_{2g}^{1}$ peak, derived from the dashed lines in (a) and (b), exhibiting the sloping regions of the crystal on the bowtie antenna with TERS spatial resolution of 15 nm.}
		\label{fig:fig2}
	\end{center}
\end{figure}

\textbf{Figure~\ref{fig:fig2}}a shows a spectral variation of tip-enhanced Raman spectroscopy (TERS) responses for a WSe$_{2}$ ML when the Au tip moves along the longitudinal axis of the bowtie antenna. The distinct TERS response of the A$_{1g}$ + E$_{2g}^{1}$ mode ($\sim$250 cm$^{-1}$, degenerate mode for out-of-plane vibration of Se atoms and in-plane vibration of W and Se atoms \cite{sahin2013, corro2014}) is observed in the nano-gap and both edges of the bowtie antenna. These three spots correspond to the slipping down regions of the transferred crystal.

In contrast, this spatial heterogeneity cannot be resolved in the plasmon-enhanced Raman profile (Figure~\ref{fig:fig2}b) due to the resolution limit. To understand the physical mechanism of the strong TERS response in the three local regions, we investigate spectral features by plotting TERS (red) and plasmon-enhanced Raman (blue) spectra in the nano-gap as shown in Figure~\ref{fig:fig2}c. While the A(M) mode ($\sim$235 cm$^{-1}$, asymmetric phonon mode at the M point \cite{terrones2014, zhang2015}) shows no distinct enhancement, the A$_{1g}$ + E$_{2g}^{1}$ mode shows significant signal enhancement in the TERS measurement. In the nano-gap, the vibration direction of the out-of-plane Raman mode (A$_{1g}$) and in-plane Raman mode (E$_{2g}^{1}$) are not matched with the polarization axis of the nano-gap plasmon in the bowtie antenna because of the rotated geometry of the 2D crystal caused by the dramatic sloping down structure. The strong enhancement of TERS intensity in the nano-gap well indicates the effect of the tip which enhances the z-polarized near-field dominantly. Therefore, we can assume that the E$_{2g}^{1}$ mode is the dominant vibrational mode in the TERS peak of the A$_{1g}$ $+$ E$_{2g}^{1}$ mode. i.e., the E$_{2g}^{1}$ mode is in parallel with respect to the tip-axis and the dominant near-field polarization \cite{tong2019, zeng2013, zhao2013}. Figure~\ref{fig:fig2}d shows TERS (green) and plasmon-enhanced Raman (blue) intensity profiles for the A$_{1g}$ $+$ E$_{2g}^{1}$ peak to quantify the spatial resolution of our approach, i.e., $<$ 15 nm.

\subsection{Radiative emission of localized excitons at room temperature}

\begin{figure}
	\begin{center}
		\includegraphics[width=16 cm]{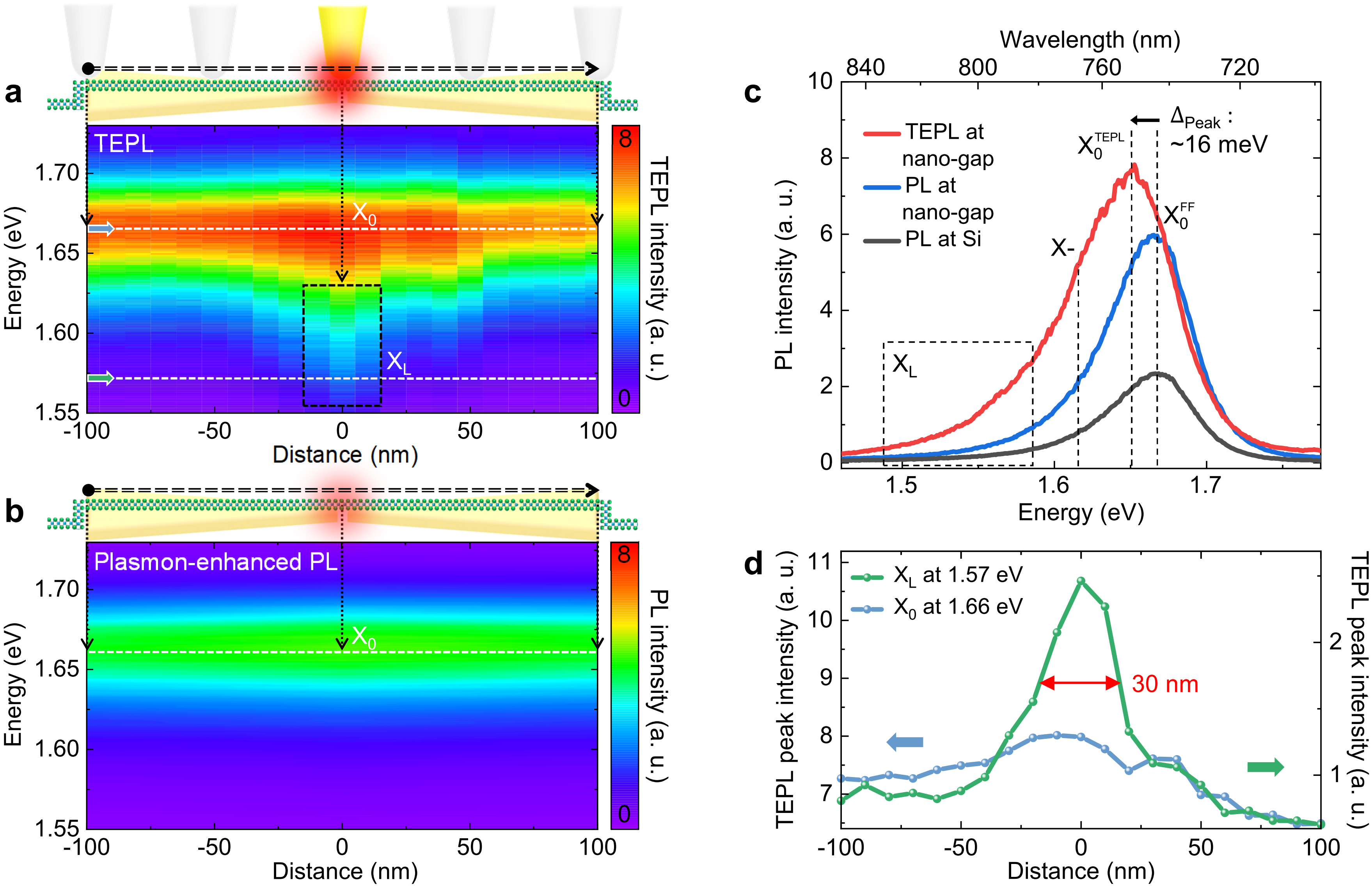}
		\caption{Probing radiative emission of localized excitons with TEPL measurement.
			Spectral TEPL (a) and plasmon-enhanced PL (b) line traces along the longitudinal center axis of the bowtie antenna with a WSe$_{2}$ monolayer measured at room temperature.
			(c) TEPL and plasmon-enhanced PL spectra of a WSe$_{2}$ monolayer at the nano-gap, derived from (a) and (b), exhibiting a red-shift of the neutral exciton peak (X$_{0}$) and the new emerging localized exciton peak (X$_{L}$) in the TEPL spectrum. Far-field PL spectrum of a WSe$_{2}$ monolayer on the Si substrate is shown for comparison.
			(d) TEPL intensity profiles of X$_{0}$ (blue) and X$_{L}$ (green) peaks, derived from the white dashed lines in (a), exhibiting the nanoscale spatial extent of radiative emission of the localized excitons.}
		\label{fig:fig3}
	\end{center}
\end{figure}

\textbf{Figure~\ref{fig:fig3}}a shows a variation of the TEPL spectra of the WSe$_{2}$ ML when the Au tip moves along the longitudinal axis of the bowtie antenna. 
When the tip is located on the Au surface of the bowtie antenna, the TEPL signal of X$_{0}$ is observed, yet with no PL response of X$_{L}$. 
In contrast, the pronounced spectral change in TEPL signal is observed when the tip is located above the nano-gap. 
As indicated in Figure~\ref{fig:fig3}a (dashed box, X$_{L}$), the additional lower-energy shoulder emerges near the nano-gap which corresponds to the radiative emission of X$_{L}$.

On the other hand, the X$_{L}$ emission is not observed in the plasmon-enhanced far-field PL spectra (Figure~\ref{fig:fig3}b) due to the diffraction limited spatial resolution. To investigate the modified spectral feature with the triple-sharp-tips structure in detail, we compare the plasmon-enhanced far-field PL and TEPL spectra in the nano-gap (derived from Figure~\ref{fig:fig3}a-b), as shown in Figure~\ref{fig:fig3}c. As for the comparison with a far-field PL spectrum of a WSe$_{2}$ ML on Si substrate (black), a significant PL enhancement of the X$_{0}$ is observed with the bowtie antenna (blue) owing to the strong in-plane field localization in the nano-gap, yet with no modification of spectral shape. In contrast, the TEPL spectrum (red) shows a peak shift of neutral exciton (X$_{0}$: from ~1.667 eV to ~1.651 eV), as well as new emerging shoulders of the localized excitons (X$_{L}$ at ~1.580 eV) and the trions (X- at ~1.637 eV). Note that we assign the X- peak from the energy difference of $\sim$30 meV compared to the PL energy of the X$_{0}$, based on previous studies \cite{bao2015, jones2013, you2015}. The observed spectral red-shift of $\sim$16 meV for the X$_{0}$ peak is attributed to the applied tensile strain in the nano-gap and corresponds to $\sim$0.3$\%$ strain based on previous studies \cite{schmidt2016}. This spectral shift cannot be observed in the plasmon-enhanced PL (Figure~\ref{fig:fig3}b) due to the limited spatial resolution in far-field measurement.

To quantify the spatial extent of bound excitons in the strain induced region, we plot the TEPL intensity profiles for the X$_{0}$ and X$_{L}$ peaks as shown in Figure~\ref{fig:fig3}d. In contrast to the gradual increase of the X$_{0}$ peak near the nano-gap, the X$_{L}$ peak shows dramatic enhancement in the vicinity of the nano-gap with a narrow full-width at half-maximum (FWHM) of $\sim$30 nm. Although the strain is formed in the 5 nm gap of the bowtie antenna, the created X$_{L}$ are naturally diffused into the crystal face \cite{park2016}. The observed length scale is in good agreement with the exciton diffusion length of a WSe$_{2}$ ML at room temperature \cite{park2016, zande2013}. Note that the observed spatial extent of $\sim$30 nm for the X$_{L}$ emission is not limited by spatial resolution of TEPL because we achieve a spatial resolution of $<$15 nm with the same tip for TERS imaging (Figure~\ref{fig:fig2}a and d).

\begin{figure}
	\begin{center}
	\includegraphics[width=16 cm]{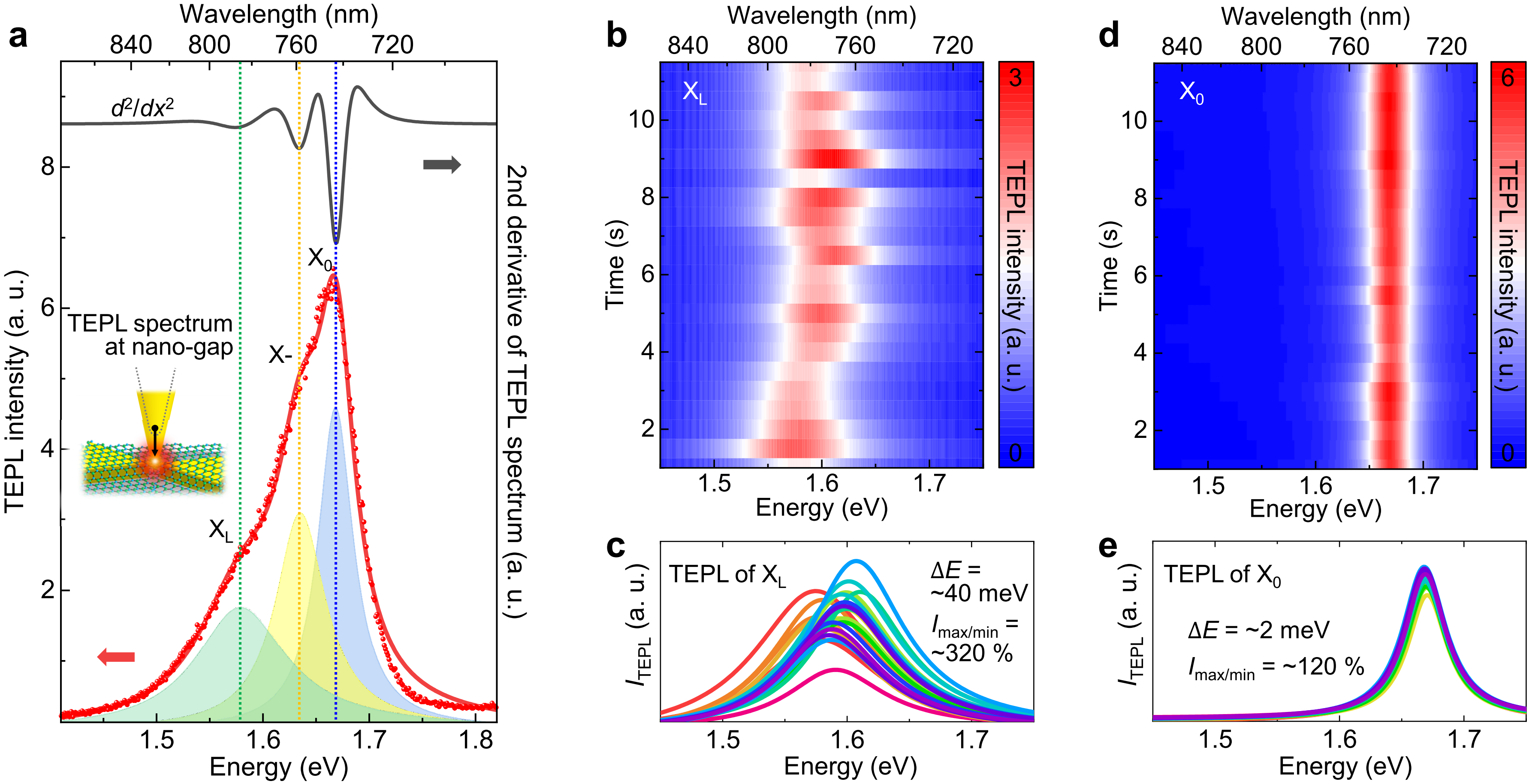}
	\caption{Inducing radiative emission of localized excitons with nano-optical antenna-tip control. 
		(a) TEPL spectrum of a WSe$_{2}$ monolayer in the triple-sharp-tips structure with a closely approached Au tip on the crystal ($<$ 1 nm) in the nano-gap. Neutral excito (X$_{0}$, blue), trion (X-, yellow), and localized exciton (X$_{L}$, green) peaks are assigned through the fitting of Voigt line shape function and reconfirmed with the second derivative of the TEPL spectrum (gray line). In the TEPL spectrum, red dots and red lines correspond to the measurement data and fit curve, respectively.
		(b, d) TEPL evolution of the X$_{L}$ and X$_{0}$ peaks with respect to time. To confirm the pronounced spectral fluctuation of the X$_{L}$ peak in comparison with the X$_{0}$ peak, all the X$_{L}$ and X$_{0}$ spectra (c, e) are derived with Voigt fitting from the time-series data.}
	\label{fig:fig4}
	\end{center}
\end{figure}

In the hyperspectral TEPL profiling (Figure~\ref{fig:fig3}a), the distance between the tip and sample is maintained at $\sim$3 nm with shear-force feedback. We then approach the Au tip further into the nano-gap to observe clearer TEPL spectrum from the triple-sharp-tips structure by increasing the field localization as well as the Purcell factor enhancement. \textbf{Figure~\ref{fig:fig4}}a shows the measured TEPL spectrum when the tip-sample distance is $<$1 nm. From fitting to the Voigt line shape function, we can derive distinct peaks of X$_{0}$ (blue), X- (yellow), and X$_{L}$ (green) responses. In addition, we calculate the second derivative of the TEPL spectrum (black) to reconfirm the peak positions from them. As can be seen in Figure.~\ref{fig:fig4}a, the minimum points of the second derivative curve exactly correspond to X$_{0}$, X-, and X$_{L}$ peaks. The reproducibility of our deterministic X$_{L}$ emission control is confirmed through the repeated experiments with different bowtie antennas (Figure S2 in Supporting Information)

We then measure the TEPL spectra as a function of time at the fixed tip-sample distance to investigate spectral fluctuation and blinking behavior of the X$_{L}$ PL. Figure~\ref{fig:fig4}b and d show TEPL evolutions of the X$_{L}$ and X$_{0}$ peaks derived from Voigt fitting (Figure S3 in Supporting Information for details). While the X$_{0}$ peaks show stable emission in intensity and energy, the X$_{L}$ peaks show distinct spectral and intensity fluctuations. All the emission spectra of X$_{L}$ and X$_{0}$ over time are shown in Figure~\ref{fig:fig4}c and e to better visualize the spectral variation of the X$_{L}$ peaks. Peak energy shifts are observed as large as $\sim$40 meV for the X$_{L}$ peaks and $\sim$2 meV for the X$_{0}$ peaks. In addition, the ratio of maximum and minimum peak intensities are $\sim$320$\%$ and $\sim$120$\%$ for the X$_{L}$ and X$_{0}$ peaks. Furthermore, the X$_{L}$ emission shows a $\sim$2.5$\times$ broader PL linewidth compared to the X$_{0}$ emission since different confinement potentials are formed across the strain-induced regions and the X$_{L}$ are trapped with having different energy states \cite{branny2017, he2015, hidouri2016}. Therefore, the observed spectral fluctuations and broad linewidth of the X$_{L}$ show the typical characteristics of single quantum emitters \cite{tonndorf2015, luo2019}.

\subsection{Electromagnetic simulations of the optical field and radiative decay rate}

\begin{figure}
	\begin{center}
		\includegraphics[width=16 cm]{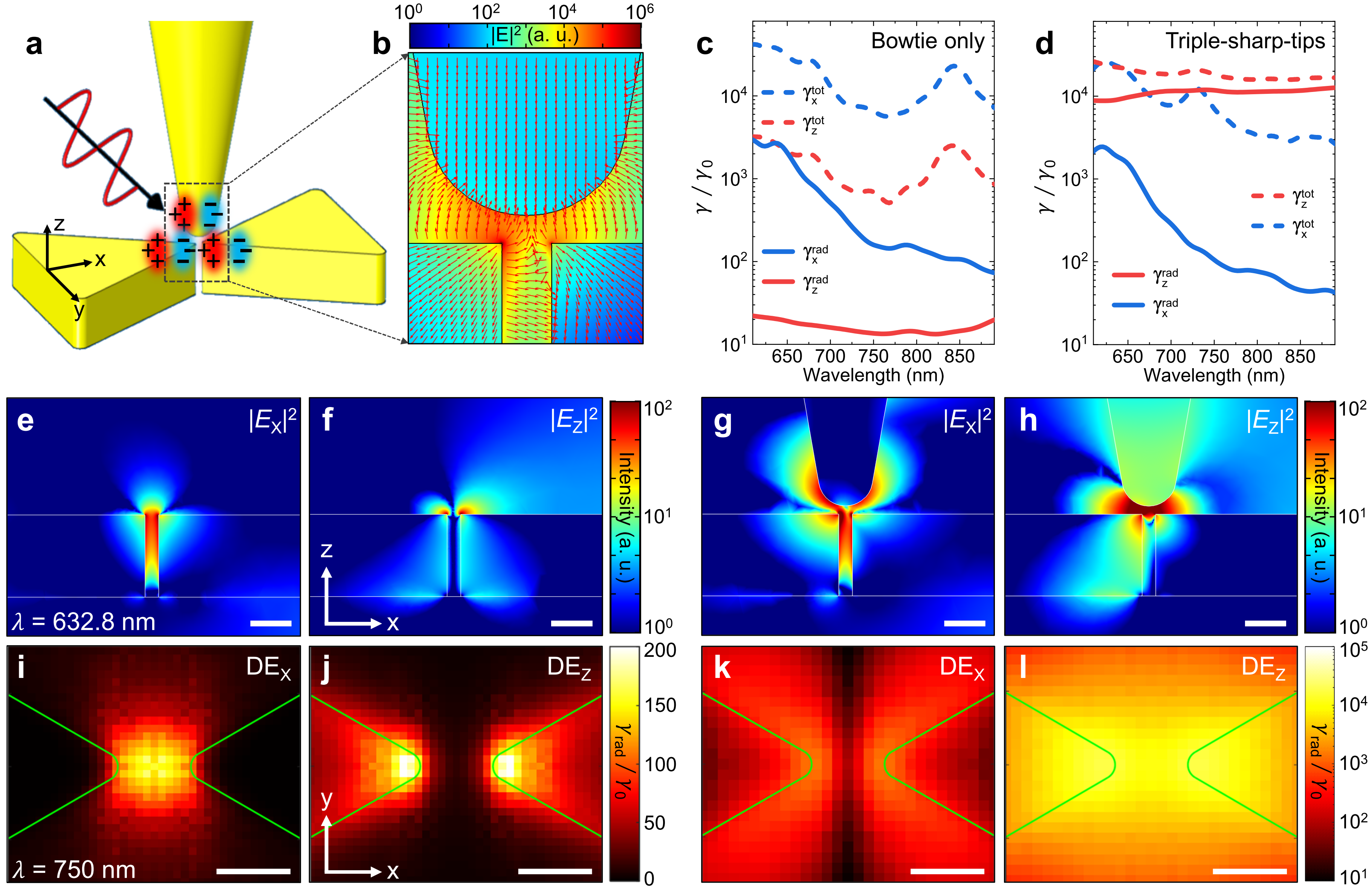}
		\caption{Electromagnetic simulations of the spatial optical properties in the triple-sharp-tips structure. 
			(a) Schematic of the triple-sharp-tips model for three-dimensional numerical analysis of the field distribution of the excitation light and the radiative-decay-rate enhancement of the emission light.
			(b) Simulated optical field intensity (${\left|E\right|}^{2}$) distribution and vector-field map in the nano-gap of the triple-tips.
			(c, d) Simulated cross-sections for the enhancement of total (dashed lines) and radiative (solid lines) decay rates in the nano-gap of the bowtie antenna (c) and triple-sharp-tips (d) structure. The cross-sections are simulated for the x- (blue) and z-polarized (red) dipole emitters placed 1.5 nm above the top surface of the bowtie antenna.
			Simulated in-plane (e, g) and out-of-plane (g, h) optical field intensity distributions in the bowtie antenna (e, f) and the triple-sharp-tips (g, h) structure. The wavelength of the excitation beam is fixed to 633 nm and scale bars are 15 nm.
			Simulated radiative-decay-rate enhancement ($\gamma^{rad}$/$\gamma_{0}$) distributions in the bowtie antenna (i, j) and  triple-sharp-tips (k, l) structure for x- (i, k) and z-polarized (j, l) dipole emitters placed 1.5 nm above the top surface of the bowtie antenna. The wavelength of emission is fixed to 750 nm and scale bars are 5 nm.}
		\label{fig:fig5}
	\end{center}
\end{figure}

To quantitatively investigate the optical properties of the triple-sharp-tips cavity, we model it (\textbf{Figure~\ref{fig:fig5}}a) and perform numerical analysis using finite-element method (FEM) for the field distribution and finite-difference time-domain (FDTD) for local density of states (LDOS) calculation (see Experimental Section for details). When an excitation optical field (632.8 nm) is coupled to the cavity, electric charges of the Au nano-structures resonantly oscillate with the distribution illustrated in Figure~\ref{fig:fig5}a. Figure~\ref{fig:fig5}b shows the simulated optical field intensity distribution of the cavity with a vector-field map to understand the excitation rate enhancement. In the cavity, the excitation field intensity is enhanced as high as $\sim$10$^{6}$ with the dominant out-of-plane optical field in the nano-gap between the etched tip and bowtie tips.

We simulate the decay-rate enhancement of a spontaneous emitter placed in the cavity (1.5 nm above the top surface of the bowtie antenna) with respect to the emission wavelength. Figure~\ref{fig:fig5}c and d show the cross-sections of the total- (dashed) and radiative-decay-rate (solid) enhancement of the x- (blue) and z-axis (red) polarized dipole emitters for the bowtie only (c) and triple-sharp-tips (d) cavities. While the nonradiative-decay-rate ($\gamma^{nonrad}$ $=$ $\gamma^{tot}$ - $\gamma^{rad}$) is quite large for both x- and z-polarized dipole emitters in the bowtie-only cavity (Figure~\ref{fig:fig5}c), the $\gamma^{nonrad}$ in the triple-sharp-tips cavity is significantly decreased for the z-polarized emitter $\gamma_{z}^{tot}$ in Figure~\ref{fig:fig5}d) due to Purcell effect in the confined cavity structure \cite{park2018}.

We then calculate the in-plane and out-of-plane optical field intensity distribution separately to compare the excitation rate for X$_0$ (${\left|E_{x}\right|}^{2}$) and X$_{L}$ (${\left|E_{z}\right|}^{2}$) in the cavities, as shown in Figure~\ref{fig:fig5}e-h. When the Au tip is added to the bowtie antenna, ${\left|E_{x}\right|}^{2}$ at 1.5 nm above the bowtie nano-gap (Fig.~\ref{fig:fig5}g) shows comparable intensity with the bowtie-only antenna (Figure~\ref{fig:fig5}e). In contrast, ${\left|E_{z}/E_{0}\right|}^{2}$ is increased from $\sim$10 for the bowtie-only antenna to $\sim$10$^{2}$ for the triple-sharp-tips antenna by the added Au tip effect (Figure~\ref{fig:fig5}h) and we can estimate the excitation rate for X$_{L}$ in our experiment.

Next, to quantify the enhanced spontaneous emission rate from the Purcell effect in our TEPL measurements, we simulate the distribution of the radiative-decay-rate enhancement ($\gamma^{rad}$/$\gamma_{0}$) for the x- and z-polarized dipole emitters (750 nm, 2D plane 1.5 nm above the top surface of the bowtie antenna) in both cavities, as shown in Figure~\ref{fig:fig5}i-l. Similar to the confirmed out-of-plane cavity effect on the excitation rate, we obtain ${\gamma^{rad}}/{\gamma_{0}}$ as high as $\sim$10$^{4}$ for the z-polarized emitter in the triple-sharp-tips cavity (Figure~\ref{fig:fig5}l). These simulated excitation and emission rates for the z-polarized emitters in the cavity explain the room temperature observation of X$_{L}$ emission in our TEPL experiment. In general, X$_{L}$ PL is hard to observe at room temperature with far-field measurement because of the rapid decrease in quantum yield with increasing temperature caused by thermal activation of carriers and low exciton binding energy of X$_{L}$ \cite{ross2013, huang2016, pandey2019}. On the other hand, the highly increased emission rate as well as the excitation field localization in the triple-sharp-tips cavity allows us to induce and probe X$_{L}$ TEPL in our experiment \cite{luo2018}.

\section{Discussions}

Our experiment is carefully designed to induce, probe, and control the radiative emission of X$_{L}$ at room temperature. We fabricate extremely sharp bowtie tips \cite{kim2020} (RoC of $<$1 nm) and transfer a WSe$_{2}$ monolayer on it. The monolayer crystal has a steep curve in the nano-gap which gives rise to an anisotropic tensile strain in the downward direction \cite{cotrufo2019}. This local strain leads to induce X$_{L}$ with a vertically oriented transition dipole moment \cite{tonndorf2015, iff2018, lundskog2014}.

Although X$_{L}$ are created in this device, their radiative emission is not observed with far-field measurement. Since the orientation of the transition dipole moment of X$_{L}$ is perpendicular to the polarization of the nano-gap plasmon of the bowtie antenna, the X$_{L}$ cannot be effectively excited, radiatively emitted, and optically detected. Furthermore, the emission cannot be selectively and locally probed with far-field optics due to the diffraction limit. Hence, to locally probe the X$_{L}$ emission, we introduce an Au tip to form triple-sharp-tips structures (Figure~\ref{fig:fig1}a).

In this nano-cavity device, we can significantly increase the excitation rate (${\left|{E_{z}}/{E_{0}}\right|}^{2}$ $=$ $\sim$10$^2$) of the vertically oriented X$_{L}$ owing to the strong out-of-plane field localization (Figure~\ref{fig:fig5}h). In addition, the spontaneous emission rate is largely enhanced by the Purcell effect (${\gamma^{rad}}/{\gamma_{0}}$ $=$ $\sim$10$^{4}$) because the mode volume in triple-sharp-tips cavity is extremely small (Figure~\ref{fig:fig5}l) compared to a conventional Au tip – metal mirror substrate platform \cite{kleemann2017, park2018}. Based on our electromagnetic simulation results, we can estimate a TEPL enhancement factor of the localized excitons as high as $\sim$10$^{6}$ with an equation given by \cite{park2018}

\begin{center}
	$<$EF$>$ $=$ ${\left|\frac{E_{z}}{E_{0}}\right|}^{2}$$\times\frac{\gamma^{rad}}{\gamma_{0}}$.
\end{center}

As a result, our triple-sharp-tips approach in combination with TEPL nano-spectroscopy enables us to induce and probe X$_{L}$ emission. Moreover, precise 3D tip positioning ($<$0.2 nm) based on atomic force control provides the ability to dynamically engineer the cavity mode volume, field strength, and corresponding X$_{L}$ emission, as can be seen in the lateral (Figure~\ref{fig:fig3}a) and vertical (Figure~\ref{fig:fig4}a) tip-positioning effects. Therefore, our work demonstrates a systematic method to induce, probe, and control the single quantum emitters in atomically thin semiconductors even at room temperature.

\section{Conclusion}

In conclusion, we have demonstrated that ultracompact tip-enhanced nano-cavity spectroscopy with a triple-sharp-tips geometry gives access to localized states of excitons, beyond the limitations of previous X$_{L}$ studies, such as observation at cryogenic temperature, limited spatial resolution, or the lack of systematic control. Most importantly, the carefully designed cavity structure provides a Purcell factor as high as $\sim$10$^4$ for X$_{L}$ emitters with the ability to dynamically control the cavity mode volume (V) and corresponding radiative emission rate ($\propto$ ${1}/{V^2}$) \cite{schuller2013, luo2017, pelton2015, luo2019}. We expect that this platform enabling bright X$_{L}$ emission and robust control at room temperature will allow a range of device applications for quantum information technology. The presented PL enhancement ($>$ 10$^6$) method will be applied to greatly improve the efficiency and performance of quantum photonic devices \cite{zwiller2004, schlehahn2016}. In addition, the previously demonstrated electrically-driven single quantum emission (electroluminescence) in 2D materials at cryogenic temperature \cite{ross2014, baugher2014, pospischil2014, luo2018} can be realized in ambient condition by adopting our cavity structure. Furthermore, our tip-enhanced nano-cavity and -spectroscopy approach can be extended to characterize other low-dimensional quantum materials, such as zero-dimensional quantum dots \cite{vahala2003, dusanowski2020} and one-dimensional quantum wires \cite{luo2017}, and induce and control their emitter-cavity interactions from the weak to the strong coupling regime \cite{park2019}.

\section{Experimental Section}
\threesubsection{Sample preparation}\\
To transfer the chemical vapor deposition (CVD)-grown WSe$_2$ monolayer onto a flat Au film/coverslip, a wet transfer process was used.
As a first step, poly(methyl methacrylate) (PMMA) was spin-coated onto WSe$_2$ monolayer grown on the SiO$_2$ substrate.
The PMMA coated WSe$_2$ monolayer was then separated from the SiO$_2$ substrate using a hydrogen fluoride solution, and carefully transferred onto the bowtie device after being rinsed in distilled water to remove residual etchant.
Then, it is dried naturally for 6 hours to improve the adhesion.
Lastly, the PMMA was removed using acetone.

\threesubsection{Bowtie fabrication}\\
The sharp bowtie nanoantenna fabrication starts on a 500 $\mu$m thick silicon substrate. Using standard electron beam lithography (Elionix ELS-7800, acceleration voltage$:$ 80 kV, beam current: 50 pA), the half bowtie patterns were defined on the copolymer (Microchem, MMA (8.5) MAA EL-8) / PMMA (Microchem, 495 PMMA A2) bilayer positive tone resist having different solubility in the developer. Such a bilayer process produces a T-shaped profile of the resist after development. The copolymer layer was spin-coated (5000 rpm, 60 seconds) and baked at 150$^{\circ}$C on the hotplate and its final thickness was about 250 nm. The PMMA layer was spin-coated (2000 rpm, 60 seconds) on the former copolymer layer and baked at 180$^{\circ}$C, its thickness was about 60 nm. After electron beam exposure, the copolymer / PMMA bilayer resist were developed in MIBK$:$ IPA 1$:$3 solutions for 23 minutes at 4$^{\circ}$C. A cold development with a longer develop time enables continuous development in the copolymer area affected by secondary electrons from the substrate while development in the PMMA layer is completed. As a result, a highly unstable T-shape resist profile can be made. The developed patterns were rinsed with IPA for 30 seconds and to dry out the remaining liquid, nitrogen gas was blown on the patterned area directly. The unstable resist patterns collapsed, and the collapsed pillars leaned on the adjacent wall where half bowtie patterns were inscribed. This kind of process, called cascade domino lithography, can realize ultra-sharp-edged photoresist masks with single digit nanometer scale gap size. The junctions between the rounded edge planes of the pillar and wall formed a sharp gap spacing resist mask. On the mask, Cr (3 nm) and Au (50 nm) were deposited by electron beam evaporation (KVT KVE-ENS4004), followed by the standard lift-off process. The fabricated bowtie antenna has a 1 nm-radius of curvature and 5 nm gap size, which cannot be realized by conventional electron beam lithography processes.

\threesubsection{TERS/TEPL setup}\\
The prepared WSe$_{2}$ ML on bowtie antenna was loaded on a piezo-electric transducer (PZT, P-611.3x, Physik Instrumente) for XYZ scanning with $<$0.2 nm positioning precision. The Au tip (apex radius of $\sim$10 nm) fabricated with a refined electrochemical etching protocol was attached to a quartz tuning fork (resonance frequency of 32.768 kHz) to regulate the distance between the tip and sample under a shear-force AFM operated by a digital AFM controller (R9$+$, RHK Technology). For TEPL and TERS experiment, a conventional optical spectroscopy set-up was combined with the home-built shear-force AFM. For a high quality wavefront of the excitation beam, a He-Ne laser (632.8 nm, $\le$ 1 mW) was coupled and passed through a single-mode fiber (core diameter of $\sim$3.5 $\mu$m) and collimated again using an aspheric lens. The collimated beam was then passed through a half-wave plate to make the excitation polarization parallel with respect to the tip axis. Finally, the beam was focused onto the Au tip using a microscope objective (NA$=$0.8, LMPLFLN100$\times$, Olympus) with a side-illumination geometry. To ensure highly efficient laser coupling to the Au tip, the tip position was controlled with $\sim$30 nm precision by using Picomotor actuators (9062-XYZ-PPP-M, Newport).  TEPL and TERS responses were collected using the same microscope objective (backscattering geometry) and passed through an edge filter (LP02-633RE-25, Semrock) to cutoff the fundamental line. The signals were then dispersed onto a spectrometer (f$=$328 mm, Kymera 328i, Andor) and imaged with a thermoelectrically cooled charge-coupled device (CCD, iDus 420, Andor) to obtain TEPL and TERS spectra. Before the experiment, the spectrometer was calibrated with a Mercury lamp. A 150g$/$mm grating with 800 nm blazed (spectral resolution of 0.62 nm) and 600g$/$mm grating with 500 nm blazed (spectral resolution of 3.75 cm$^{-1}$) were used for PL and Raman measurements, respectively. 

\threesubsection{Numerical analysis}\\
Three dimensional full-wave simulations were performed to numerically analyze field enhancement and local density of states (LDOS) near the bowtie and the tip. Field enhancements (Figure~\ref{fig:fig5}b, e–h) were calculated using the Wave Optics module of COMSOL Multiphysics. A TM-polarized planewave incidence at 45$^\circ$ from air to a single bowtie with the tip was considered using the scattered field formulation surrounded by perfectly matched layer (or PML). LDOS (Figure~\ref{fig:fig5}c, d, i–l) were calculated using Lumerical FDTD. Total radiated power from an electric dipole and absorbed power to the total system were calculated to obtain the total and nonradiative decay rates \cite{mack2017}. Optical properties of silicon were taken from Palik \cite{palik1985}, and gold from Johnson and Christy \cite{johnson1972}.

\medskip
\textbf{Supporting Information} 
Supporting Information is available from the Wiley Online Library or from the author.

\medskip
\textbf{Acknowledgements} 
This work was supported by the National Research Foundation of Korea (NRF) grants (No. 2019K2A9A1A06099937, and 2020R1C1C1011301). J.R. acknowledges the NRF grants (NRF-2019R1A2C3003129, CAMM-2019M3A6B3030637, NRF-2019R1A5A8080290, NRF-2018M3D1A1058998) funded by the Ministry of Science and ICT (MSIT). I.K. aknowledges the NRF Global Ph.D. fellowship (NRF-2016H1A2A1906519) funded by the Ministry of Education of the Korea govermnent. Y.K. acknowledges a fellowship from the Hyundai Motor Chung Mong-Koo Foundation. M.S.J. thanks the Creative Materials Discovery Program through NRF funded by the MSIT (NRF-2019M3D1A1078304). The authors thank D.S.L. and J.H.C. for the assistance of figure illustraion.

\medskip

%
\bibliographystyle{MSP}
\bibliography{bowtie}


\end{document}